\documentclass{article}
\usepackage{fancyhdr}
\usepackage[a4paper]{geometry}
\usepackage{graphicx}
\usepackage{xcolor}
\usepackage[utf8]{inputenc} % allow utf-8 input
\usepackage[T1]{fontenc}    % use 8-bit T1 fonts
\usepackage{hyperref}       % hyperlinks
\usepackage{url}            % simple URL typesetting
\usepackage{booktabs}       % professional-quality tables
\usepackage{amsfonts}       % blackboard math symbols
\usepackage{nicefrac}       % compact symbols for 1/2, etc.
\usepackage{microtype}      % microtypography
\usepackage{lipsum}
\usepackage[misc]{ifsym}
\usepackage[backend=biber, sorting=none, style=nature, citestyle=numeric-comp]{biblatex}
\usepackage[labelfont=bf]{caption}

\title{\textbf{Deep Learning Identifies Neuroimaging Signatures of Alzheimer's Disease Using Structural and \\Synthesized Functional MRI Data}}

\makeatletter
\def\@fnsymbol#1{\ensuremath{\ifcase#1\or \dagger\or 1 \or 2 \or 3 \or 4 \or 5 \or \hbox{\Letter} \or 6 \else\@ctrerr\fi}}
\makeatother

\author{Nanyan Zhu \thanks{These authors contributed equally to this work and are joint first-authors.} \hspace{1pt} \thanks{Department of Biological Sciences and the Taub Institute, Columbia University, New York, NY, USA.}
\and
Chen Liu \footnotemark[1] \hspace{1pt} \thanks{Department of Electrical Engineering and the Taub Institute, Columbia University, New York, NY, USA.}
\and
Xinyang Feng \thanks{Department of Biomedical Engineering, Columbia University, New York, NY, USA.}
\and
Dipika Sikka \footnotemark[4]
\and
Sabrina Gjerswold-Selleck \footnotemark[4]
\and
Scott A. Small \thanks{Department of Neurology, the Taub Institute, the Sergievsky Center, Radiology and Psychiatry, Columbia University, New York, NY, USA.}
\and
Jia Guo \thanks{Department of Psychiatry, Mortimer B. Zuckerman Mind Brain Behavior Institute, Columbia University, New York, NY, USA.} \hspace{2pt} \thanks{Email: \url{jg3400@columbia.edu}}
\and
\\for the Alzheimer's Disease Neuroimaging Initiative \thanks{Data used in preparation of this article were partially obtained from the Alzheimer’s Disease Neuroimaging Initiative (ADNI) database (\url{adni.loni.usc.edu}). As such, the investigators within the ADNI contributed to the design and implementation of ADNI and/or provided data but did not participate in analysis or writing of this report. A complete listing of ADNI investigators can be found at: \url{http://adni.loni.usc.edu/wp-content/uploads/how_to_apply/ADNI_Acknowledgement_List.pdf}}
}
\date{}

\fancypagestyle{first_page}{
  \lhead{\scriptsize© 2021 IEEE. Personal use of this material is permitted. Permission from IEEE must be obtained for all other uses, in any current or future media, including reprinting/republishing this material for advertising or promotional purposes, creating new collective works, for resale or redistribution to servers or lists, or reuse of any copyrighted component of this work in other works.
  \\\textcolor{white}{.}
  \\DOI: \url{10.1109/ISBI48211.2021.9433808} \hspace{4pt} URL: \url{https://ieeexplore.ieee.org/document/9433808}
  }
}

\begin{document}
\maketitle
\thispagestyle{first_page}

\noindent\textbf{Current neuroimaging techniques provide paths to investigate the structure and function of the brain \textit{in vivo} and have made great advances in understanding Alzheimer's disease (AD). However, the group-level analyses prevalently used for investigation and understanding of the disease are not applicable for diagnosis of individuals. More recently, deep learning, which can efficiently analyze large-scale complex patterns in 3D brain images, has helped pave the way for computer-aided individual diagnosis by providing accurate and automated disease classification. Great progress has been made in classifying AD with deep learning models developed upon increasingly available structural MRI data. The lack of scale-matched functional neuroimaging data prevents such models from being further improved by observing functional changes in pathophysiology. Here we propose a potential solution by first learning a structural-to-functional transformation in brain MRI, and further synthesizing spatially matched functional images from large-scale structural scans. We evaluated our approach by building computational models to discriminate patients with AD from healthy normal subjects and demonstrated a performance boost after combining the structural and synthesized functional brain images into the same model. Furthermore, our regional analyses identified the temporal lobe to be the most predictive structural-region and the parieto-occipital lobe to be the most predictive functional-region of our model, which are both in concordance with previous group-level neuroimaging findings. Together, we demonstrate the potential of deep learning with large-scale structural and synthesized functional MRI to impact AD classification and to identify AD's neuroimaging signatures.
}

\section*{Introduction}

\begin{figure}[!b]
\centering
\includegraphics[width=450pt,trim=10 10 10 10,clip]{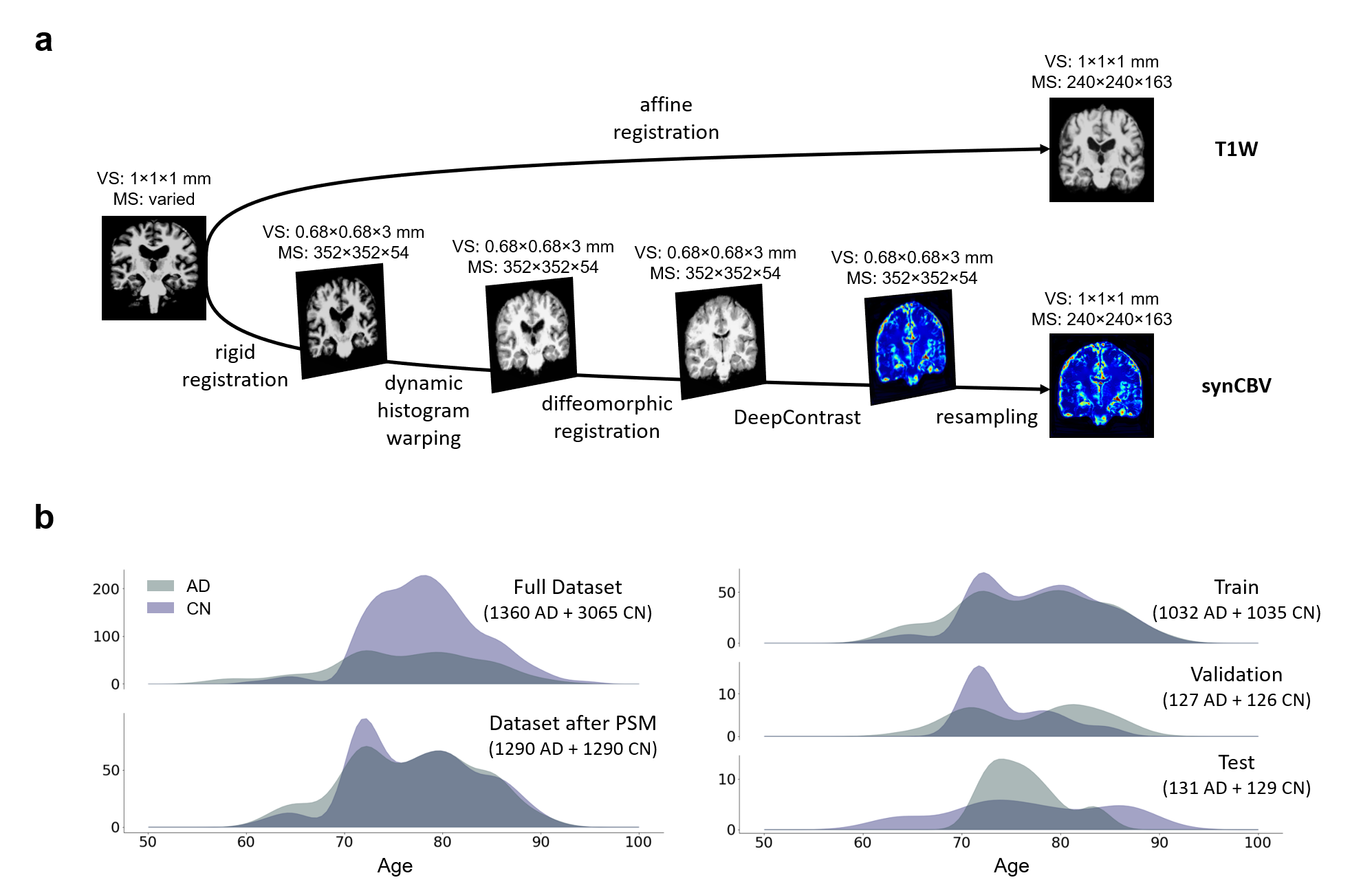}
\caption{\textbf{Data overview and partitioning.}
\textbf{a.}~The data processing pipeline that yielded the inputs to the classification models. VS: voxel size; MS: matrix size (of the entire scan). \textbf{b.}~Left: Age distributions of the subjects in the entire dataset~(top) and in the subset after propensity-score matching of age~(bottom). Right: Age distributions of the subjects assigned to the train, validation and test sets. AD: Alzheimer's diseased; CN: cognitive normal.}
\label{data_distribution}
\end{figure}

Alzheimer’s Disease~(AD) is a progressive neurodegenerative disease marked by beta amyloid plaques and neurofibrillary tangles made of hyperphosphorylated tau protein. More than five million Americans are living with AD in 2020, and AD is the sixth leading cause of death in the United States~\cite{AD2020}.

As a result of recent advancements in neuroimaging paired with rapid growth in deep learning technologies~\cite{Deep_learning, Deep_learning_in_healthcare}, numerous researchers have preformed AD diagnosis with computational methods utilizing noninvasive medical imaging data. Neuroimaging modalities can be generally categorized into two classes: structural and functional. Structural imaging delineates the anatomy and morphometry, while functional imaging captures the underlying metabolism. In the progression of AD, functional abnormalities occur prior to structural deformations~\cite{Functional_before_structural}. It is believed as a result that functional modalities may provide more value in AD classification tasks. The utility of functional imaging in AD diagnosis is highlighted in a 2019 review~\cite{Deep_learning_AD_review} that screened for 389 research articles and showed that the best AD classification accuracy achieved were 0.914 using structural data~(T1W MRI)~\cite{Best_structural_2019}, 0.96 using functional data~(FDG- and Amyloid-PET)~\cite{Best_functional_2019}, and 0.988 using structural and functional data combined~(T1W MRI, FDG-PET and CSF biomarkers)~\cite{Best_structural_functional_2019}.

Despite their success, there are two major shortcomings in previous studies. First, very few of them took care of the age distribution mismatch between the AD and control cohorts, which may introduce confounding aging effects. Since in most public datasets the AD patients are generally older than the controls, instead of classifying ``AD'' versus ``cognitively-normal~(CN)'' they may be effectively classifying ``AD elderly'' versus ``CN youngsters''. The aging effects, which are by themselves classifiable from neuroimages~\cite{Aging_without_AD, Brain_age_1, Brain_age_2}, may provide additional discriminative power in such studies, inflating the classification scores.

Second, the sample sizes of these studies are generally limited, ranging from below 100 to slightly above 500 scans from AD and CN cohorts combined. While there are larger-scaled structural studies, such as one published in 2020 and thus not included in the previous review paper, where 4691 T1W MRI scans were utilized to achieve a 0.954 accuracy~\cite{Xinyang_ADscore}, to the best of our knowledge, there does not exist a similar study on large-scale functional data.

Here we attempt to address both challenges in a proof-of-concept study based on T1W MRIs from ADNI. First, we organized a large-scale structural MRI dataset with over 2500 scans from age-matched AD and CN subjects, to control for aging confound effects in deep-learning-based AD classification. Second, we took advantage of our previous research~\cite{DeepContrast, DeepC_mouse} to synthesize high-resolution functional mappings from these structural MRI, resulting in a large-scale, paired, functional dataset. We used those structural and deep-learning-synthesized functional brain images to extend a prior state-of-the-art structural-based AD classification study~\cite{Xinyang_ADscore}. We investigated several methods for modality fusion, demonstrated that the deep-learning-synthesized functional data could provide useful information for AD classification, and visualized the disease signature as identified by the best-performing classifier.

\section*{Methods}

\begin{figure}[!b]
\centering
\includegraphics[width=450pt,trim=10 10 10 10,clip]{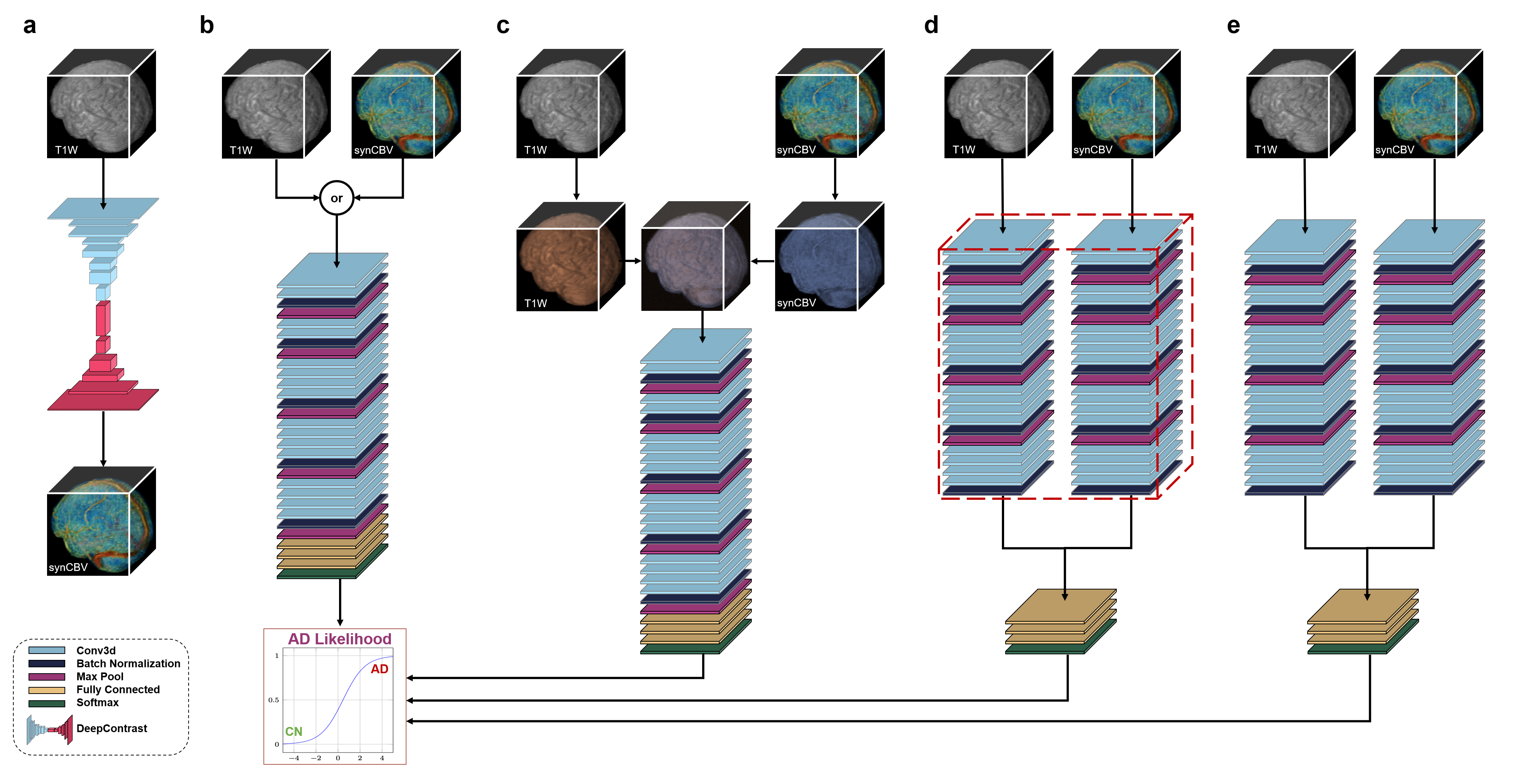}
\caption{\textbf{Network architectures implemented for AD classification.} \textbf{a.}~DeepContrast is used to generate the synthesized CBV maps from T1W scans. \textbf{b.}~VGG-19 with Batch Normalization. Used for cases with one single input modality. \textbf{c-e.}~Modified architectures for dual-modality input. \textbf{c.}~Option 1: Channel-wise combination of the two modalities. \textbf{d.}~Option 2: Modality-specific VGG encoders, with the weights shared across the two encoders. \textbf{e.}~Option 3: Modality-specific VGG encoders, but different weights across the two encoders.}
\label{AD_classification_architectures}
\end{figure}

\subsection*{Study Design}
We utilized a pre-trained network called DeepContrast~\cite{DeepContrast}. It is a deep learning approach proposed to perform quantitative structural-to-functional mapping, extracting the hemodynamic information from structural MRI. It takes in T1W MRI scans and generates voxel-level predictions of contrast uptake simulating the effects of gadolinium-based contrast agents. After normalization, the model outputs become a close approximation of cerebral blood volume~(CBV) maps, a functional imaging modality depicting basal metabolism and often used to localize functional deficits~\cite{CBV_utility_1990, CBV_utility_1991, CBV_utility_2002, CBV_utility_2009, CBV_utility_2014, CBV_utility_2014_another, CBV_utility_2016, CBV_utility_2017, CBV_utility_2020}.

We applied the DeepContrast model on a 2580-scan T1W MRI cohort and yielded 2580 synthesized CBV~(synCBV) scans, each corresponding to one T1W MRI scan.

Then, we trained multiple networks based on VGG-19 to perform the AD vs. CN binary classification task. We altered the network input with different options, including T1W MRI, synthesized CBV, or the combination of the two. We reported the respective network performances and illustrated the specific brain regions that contribute the most to the classification results.

\subsection*{Data Preprocessing and Partitioning}
We screened T1W MRI scans previously downloaded from the Alzheimer's Disease Neuroimaging Initiative~(ADNI)~\cite{ADNI} dataset, and excluded all scans except for 3~Tesla MP-RAGE acquisitions~(Fig.\ref{data_distribution}b top left). We further performed propensity score matching~(PSM)~\cite{PSM} to match the age distributions and eventually resulted in a dataset with 1290 scans of AD patients and 1290 scans of age-matched CN volunteers~(Fig.\ref{data_distribution}b bottom left).

These scans, due to prior efforts~\cite{Xinyang_ADscore}, were already skull-stripped and registered to the MNI-152 template~\cite{MNI152} using affine deformations. These scans in affine-MNI space were used as the T1W MRI dataset in our study~(Fig.\ref{data_distribution}a top).

On a separate path, we (1)~registered the scans from their native space to the unbiased template built from the data used to optimize the DeepContrast structural-to-functional mapping model, (2)~mitigated appearance difference using dynamic histogram warping~\cite{DHW_medical_images}, (3)~diffeomorphically registered them to the unbiased template, (4)~used DeepContrast followed by post-normalization to generate the synthesized CBV, and (5)~resampled to match the resolution of the prepared T1W input. Necessity and rationale for these steps were previously described~\cite{DeepContrast}. The entire pipeline is illustrated in the bottom path in Fig.\ref{data_distribution}a.

The T1W MRI scans were affine-registered to reduce variance in features such as the brain volume while still preserving differences in local anatomy which may presumably reflect AD-related effects on brain structures. The synthesized CBV scans were diffeomorphically-registered to minimize effects from structural differences.

The prepared cohort with a total of 2580 T1W scans and 2580 synthesized CBV scans, were randomly assigned to train, validation, and test sets for the AD-classification task at an 8:1:1 ratio. Randomization was performed on the subject level to prevent data leakage. AD and CN subjects were independently randomized to balance the presence of both classes in each set. The data distribution is summarized in the right half of Fig.~\ref{data_distribution}b.

\subsection*{Classification Network Architecture and Implementation}

For the AD classification tasks with one single input modality, the architecture ``VGG-19 with batch normalization''~(VGG-19BN)~\cite{VGG19BN} was used~(Fig.~\ref{AD_classification_architectures}b). When both T1W and synthesized CBV were used as inputs, each as a three-dimensional~(3D) volume, three options were experimented for information fusion. One is appending the two 3D volumes in an additional fourth axis, treating them as separate channels~(i.e., just like the R/G/B channels of color images), as illustrated in Fig.~\ref{AD_classification_architectures}c. In the last two options we used separate VGG encoders for each volume and later appended the extracted feature vectors together before entering the fully-connected layers. The two encoders may either share identical weights~(Fig.~\ref{AD_classification_architectures}d) or keep different weights~(Fig.~\ref{AD_classification_architectures}e). For any of these architectures, the input is the relevant 3D scan(s) while the output is a continuous-valued number representing the predicted AD-likelihood.

\begin{table}[!tb]
\centering
\caption{\textbf{Classification performances of the five candidates.} Sensitivity and specificity are calculated at the operating point. Accuracy at the operating point and the maximum accuracy achievable by changing the binarization threshold are respectively calculated for each candidate. ROC~AUC: area under the receiver-operating characteristics curve.}
\includegraphics[width=420pt,trim=10 10 10 10,clip]{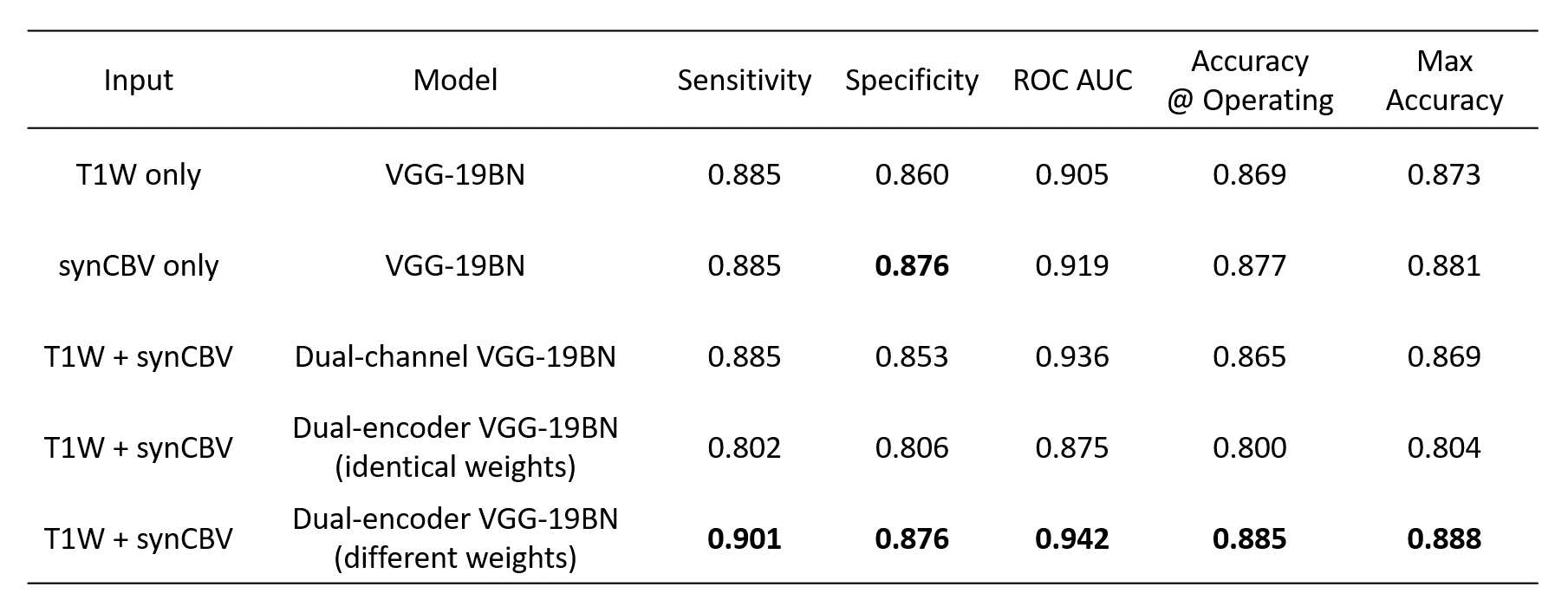}
\label{Results_table}
\end{table}

The training time is approximately 7 hours for each classifier, on a GeForce RTX2080 Ti graphics card.

To evaluate the descriptiveness of the predicted AD-likelihoods, we conducted receiver-operating characteristics~(ROC) studies to analyze the concordance between the model-generated classification and the ground truth AD/CN labels. The ROC curves, one for each trained classification model, represent the classification performances at each potential numerical threshold to binarize the predicted AD-likelihood score. The sensitivity and specificity (the sum of whom peaks at the operating point), as well as the total area under the ROC curve, demonstrate the effectiveness of the classification method. The significance of the differences among these ROC curves are calculated using DeLong's test~\cite{DeLong}.

\subsection*{Evaluation and Localization}

Further, we investigated the brain regions that had the most contributions to the AD classification task by visualizing the class activation maps~(CAM)~\cite{Class_activation_map}. We used all 131 T1W and 131 SynCBV scans from AD patients to generate an averaged CAM for each input type. We were interested in whether or not the brain regions the classifier found most relevant to the AD class were in fact physiologically meaningful.

\section*{Results}

After training the networks, we tested the five aforementioned candidates on the same stand-alone set of 131 AD scans and 129 CN scans. Classification performance using the synthesized functional data~(synCBV) alone is equal or better in every aspect than that using the structural data~(T1W) alone. Utilization of both modalities using channel combination yielded better ROC but worse accuracy compared to using any one single modality. The dual-encoder approach for modality fusion, with the encoders sharing identical weights yielded the worst performance among all candidates, whereas the same approach with the weights not shared between the encoders resulted in the best performances across all metrics. The quantitative performance metrics are summarized in Table~\ref{Results_table}.

\begin{figure}[!b]
\centering
\includegraphics[width=450pt,trim=10 10 10 10,clip]{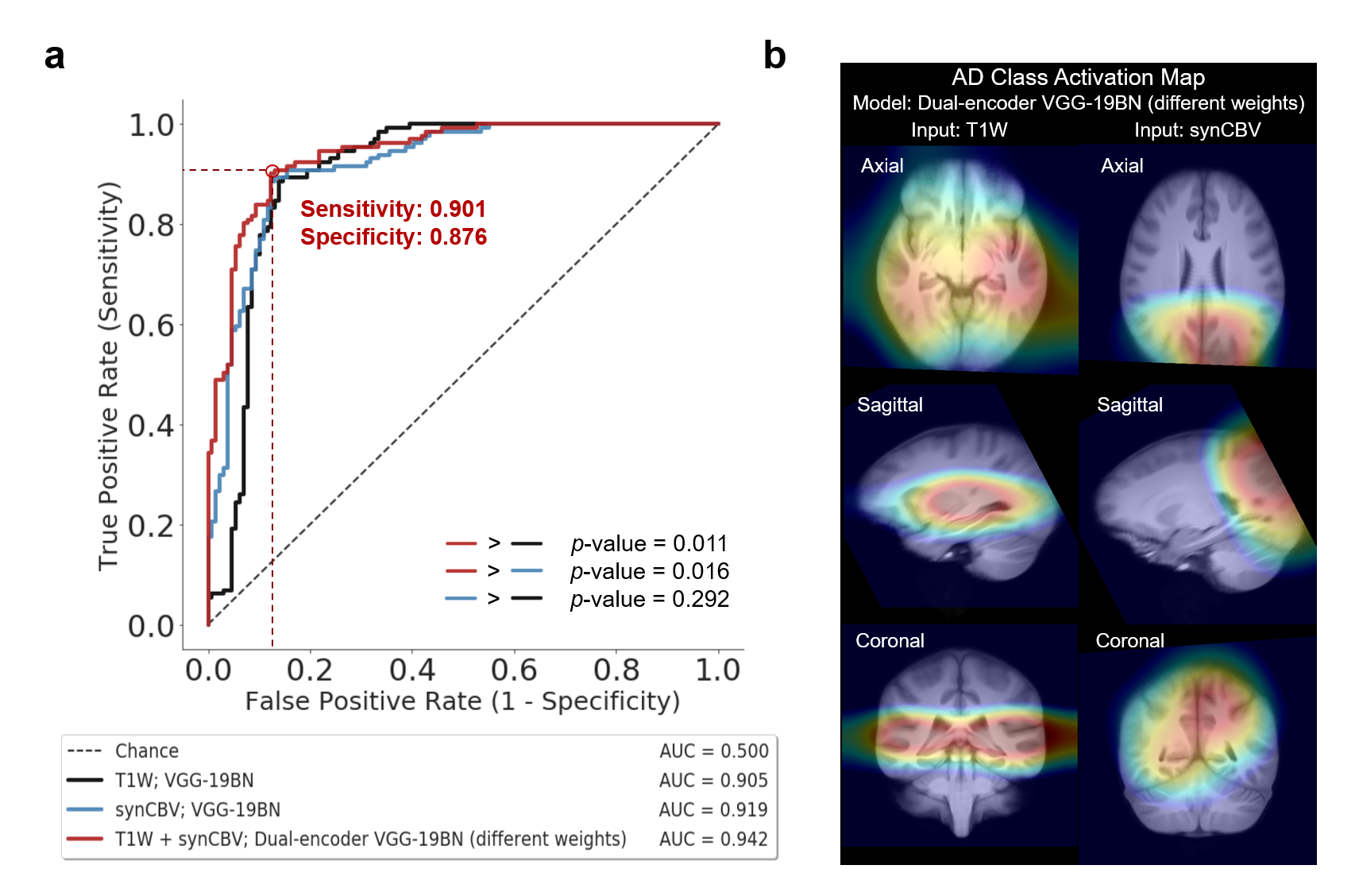}
\caption{\textbf{Receiver Operating Characteristics~(ROC) and Class Activation Maps~(CAM).} \textbf{a.}~ROC curves for the classification models using structural data only~(black), using synthesized functional data only~(blue), or using both with the ``dual-encoder with different weights" approach~(red). \textbf{b.}~The class-average CAM, calculated from all 131 AD scans, of the best performing model in response to structural and synthesized functional data.}
\label{Results_visual}
\end{figure}

When inspecting the ROC curves~(Fig.~\ref{Results_visual}a), the same trend is preserved: at each given level of true positive rate or false positive rate, in general the classifier utilizing both structural and synthesized functional data~(dual-encoder with different weights) outperforms the one with synthesized functional data only, which is slightly better than the one with structural data as the input. The result of DeLong's test indicated that the dual-encoder with different weights is significantly better than the other two candidates.

The class activation map of the best-performing classifier is illustrated in Fig.~\ref{Results_visual}b. The most discriminative structural information comes from the temporal lobe, and the most discriminative synthesized functional information comes from the parieto-occipital lobe.

\section*{Discussion}

\subsection*{Performances of the Five Networks}

Between the two single-modality networks, the one trained on the synthesized functional images yielded better AD classification compared to its counterpart trained on structural scans. Among the three dual-modality network variants, the one utilizing two encoders with separate weights achieved the highest scores across all metrics we measured. It superseded the performances of not only the other two counterparts but also the two single-modality classifiers. Training each encoder to focus on one modality independently allowed the most efficient grasp of the data distribution and eventually resulted in the best classification performance.

Our experimental findings support prior knowledge that functional abnormalities and structural deformations offer independent contributions to the task of AD classification and that the former provide more discriminative power. Taking advantage of the unique information offered by each modality, we also developed a method to constructively combine these two modalities for even better performance compared to classification with each component individually. More significantly, we showcased DeepContrast’s ability to essentially fill the void of large-scale coherent functional data, which are often sparser and less abundant than structural data, at little additional cost.

\subsection*{Interpretation of the Class Activation Maps}

The class activation map for the best-performing classifier reveals an interesting pattern of collaboration between the two encoders, each dedicated to a single input modality. The medial temporal lobe provided the most important structural information as reflected by our structural-encoder. This result is consistent with previous studies which have indicated that medial temporal atrophy is indicative of AD, and qualitative assessments of the region can be used to predict patients at risk of AD~\cite{AD_MTL_Structural_1984, AD_MTL_Structural_2002, AD_MTL_Structural_2004, AD_MTL_Structural_2006, AD_MTL_Structural_2014}. On the other hand, activation of the parietal and occipital lobes was representative of regions experiencing the most functional changes in the AD brain according to our functional-encoder, which is consistent with findings such as decreased resting state neural activity~\cite{AD_Functional_RestingState_2007, AD_Functional_RestingState_2016} and accumulated tau~\cite{AD_Functional_Tau_2017} in the parieto-occipital cortex, both of which are characteristic of AD progression.

\subsection*{Limitations and Future Work}
While our current results demonstrate great promise for the utilization of deep-learning-synthesized functional data for AD classification, the best candidate in this study has not reached the specificity and sensitivity required for clinical diagnostic biomarkers. The reasons come in two folds: First, we intentionally mitigated the confounding aging effects by curating the dataset and resulted in a less-inflated classification performance; Second, as a proof-of-concept study we have not utilized as much data as is publicly available online.

In our future work, we will better handle these two issues in the hope of presenting a more comprehensive study. First, instead of the PSM approach that would necessarily discard a certain fraction of the valuable data, we will consider other techniques that take advantage of smarter training strategies to handle class imbalance, such as class-specific augmentation or hard-example mining. Second, we will aim to utilize several other public databases such as OASIS~\cite{OASIS} and AIBL~\cite{AIBL}, aiming at enlarging our dataset to beyond 10,000 structural scans.

Moreover, with the preliminary success in AD classification, we will further apply our network architecture and methodology to the classification of other brain disorders such as Parkinson's Disease and Schizophrenia.

\subsection*{Closing Remarks: What Modalities Can be Synthesized}
Not all functional imaging modalities can be valid candidates for synthesis from structural data. CBV-fMRI used in our case has two distinct properties that render it a viable option. First, it maps the basal metabolism of the brain rather than metabolism fluctuations in response to tasks or stimulations. It makes sense to assume a one-to-one correspondence between a high resolution structural brain image and its underlying basal metabolism at resting state of the same subject. Second, the information carried by CBV-fMRI is intrinsically present in its structural counterpart~(T1W MRI), since T1W images are able to depict a certain level of blood-tissue contrast which CBV-fMRI strengthened in a selective manner. As such, the structural-to-functional mapping is meaningful and feasible from a physical standpoint.

On the other hand, if a targeted functional imaging modality depicts physical processes not captured by the structural counterpart (e.g., a structural scan acquired at resting state followed by a functional scan with visual stimulation), or if the functional information is believed to be absent from the structural modality, the structural-to-functional mapping is likely unreliable and shall undergo careful scrutiny and further rationalization before deployment.

\section*{Compliance with Ethical Standards}
This research study was conducted retrospectively using human subject data made available in open access by ADNI~\cite{ADNI}. Ethical approval was not required as confirmed by the license attached with the open access data.

\section*{Acknowledgments}
This study was funded by the Seed Grant Program and Technical Development Grant Program at the Columbia MR Research Center. This study was performed at the Zuckerman Mind Brain Behavior Institute MRI Platform, a shared resource. Data collection and sharing for this project was partially funded by the Alzheimer’s Disease Neuroimaging Initiative~(ADNI).

\section*{Conflicts of Interest}
The authors declare the following competing interests. S.A.S. serves on the scientific advisory board of Meira GTX, recently came off the scientific advisory board of Denali Theraputics, and is an equity holder in Imij Technologies. X.F., S.A.S. and J.G. have either granted patents or applications in neuroimaging for which no royalties are received.

\small \printbibliography

\end{document}